# Intertemporal Substitutability, Risk aversion and Asset Prices

Dominique Pépin
*Centre de Recherche sur l'Intégration Economique et Financière*

## Abstract

Is the elasticity of intertemporal substitution (EIS) more or less than one? This question can be answered by confronting theoretical results of asset pricing models with investor behaviour during episodes of stock market panic. If we consider these episodes as periods of high risk aversion, then lower asset prices are in fact associated with higher risk aversion. However, according to theoretical models, risky asset price is an increasing function of the coefficient of risk aversion only if the EIS exceeds unity. It may therefore be concluded that the EIS must be more than one to reconcile theory with the observed stock price decline during periods of panic.

Intertemporal Substituability, Risk aversion and Asset Prices

Pepin Dominique
*Centre de Recherche sur l'Intégration Economique et Financière*

## *Abstract*

Is the elasticity of intertemporal substitution (EIS) more or less than one? This question can be answered by confronting theoretical results of asset pricing models with investor behaviour during episodes of stock market panic. If we consider these episodes as periods of high risk aversion, then lower asset prices are in fact associated with higher risk aversion. However, according to theoretical models, risky asset price is an increasing function of the coefficient of risk aversion only if the EIS exceeds unity. It may therefore be concluded that the EIS must be more than one to reconcile theory with the observed stock price decline during periods of panic.

# 1. Introduction

Risk aversion and elasticity of intertemporal substitution (EIS) are key parameters of the individual behaviour in an intertemporal risky environment. Whereas there is a huge amount of evidence pointing towards a high risk aversion parameter[1], there is no consensus about the EIS value. According to Donaldson and Mehra (2008, p. 50), "there is no prevailing consensus estimate of this quantity, even as regards to it being greater, equal to, or less than one". Some articles support the hypothesis of a low EIS (see for example Hall 1988, Campbell and Mankiw 1989, Barsky *et al.* 1997, Ogaki and Reinhart 1998, Campbell 2003 and Yogo 2004), whereas others support the hypothesis of a high EIS (Hansen and Singleton 1982, Attanasio and Weber 1989, Vissing-Jorgensen 2002, Bansal *et al.* 2007 and Gruber 2013). Havranek (2014) examines 2,375 estimates of the EIS reported in 169 published studies. He reports that the mean estimate of the EIS is about 0.5, which is rather in favor of the low EIS hypothesis. Yet, a high value for the EIS is a precondition that the long-run risk model of Bansal and Yaron (2004) - a successful model for explaining the risk premium and other key asset markets phenomena - has to satisfy. When working with this type of model, we should assume that the EIS is more than one, which conflicts with most empirical evidence. But is the EIS actually more than one?

By confronting theoretical implications of asset pricing models to investor behaviour during episodes of stock market panic, this paper provides important additional evidence in favor of high EIS value. Firstly, observation of financial markets suggests that asset prices are low when agents are more risk adverse, and *vice versa*. Secondly, by disentangling risk aversion from intertemporal substitution in a theoretical asset pricing model, we can infer that risky asset price is an increasing function of the risk aversion parameter only if the EIS is more than one. Hence we may deduce that this last condition must be verified. If not, the risky asset price would be high when agents are more risk adverse, in particular during episodes of stock market panic.

# 2. Episodes of panic and asset prices

As individual reactions during episodes of panic are very strong, it is easier to identify the main characteristics of their preference parameters from observation of these episodes. According to financial market historians, as panic engulfs investors, the demand for risky assets decreases, pushing down asset prices (see for example Galbraith 1997). Moreover, it may be suggested that this is a strict rule: asset prices always fall during episodes of stock market panic.

From a theoretical perspective, panic (or fear) and high risk aversion cannot be disentangled in the basic model of the rational agent in a risky intertemporal context. Standard models of rational behavior depend on too few parameters to permit any distinction between fear and risk aversion.

From an empirical perspective, Lerner and Keltner (2001) have found, in a correlation study, that the more fearful individuals are less willing to take risks in a hypothetical choice situation. According to their results, fear may be an important factor determining their risk aversion level. Cohn *et al*. (2014) prime financial professionals with either a boom or a bust scenario and measure their risk aversion in two experimental investment tasks with real monetary stakes. They find that subjects who were primed with a financial bust were substantially more risk averse than those who were primed with a boom. They also claim that financial professionals are more fearful in bust condition than in boom condition, and their fear is negatively related to investments in risky assets. By administering a questionnaire to customers of an Italian bank in 2007, before the 2008 financial crisis, and in 2009, after the crisis, and by analysing the responses, Guiso *et al.* (2013) show that the subjective willingness to take risks is lower during a recession. Customers reported a lower certainty equivalent for a hypothetical lottery following the 2008 financial crisis. The conclusion which emerges from these studies is that individuals behave as though they were more risk adverse when they experience feelings of fear, which is the case during periods of stock market panic. We can therefore liken episodes of stock market panic to periods of high risk aversion, and as a conclusion, we can infer that lower asset prices result from higher risk aversion.

---

[1] A high value of the risk aversion parameter is required to explain the high risk premium, (Mehra and Prescott 1985).

Such a conclusion may seem obvious, but in fact the theoretical conclusion from asset pricing models is that higher risk aversion does not always imply lower asset prices, which is perhaps somewhat counterintuitive. According to these models, asset prices are lower when agents are more risk adverse only if preference parameters are restricted in a precise way: the EIS must be more than one, as will be demonstrated in the next section.

### 3. The analytical framework

Consider the same environment as the one described by Lucas (1978), Mehra and Prescott (1985), Epstein (1988) and Weil (1989). A perishable consumption good, a fruit, is produced by non-reproducible identical trees whose number is normalised to one. Let $q_t$ denote the dividend (the number of fruit falling from the tree) collected at time t, associated with holding the single equity share. It is assumed that the production growth, $y_{t+1} = q_{t+1}/q_t$, follows an i.i.d. lognormal process :

$$\ln y_{t+1} = \Delta \ln q_{t+1} \sim \text{i.i.d. } N(\mu, \sigma^2). \tag{1}$$

The agent's preferences are described by a recursive utility function of Epstein and Zin (1989) and Weil (1989). The life-time utility $U_t$ of the agent satisfies:

$$U_t = \left[ (1-\beta) c_t^{1-\frac{1}{\psi}} + \beta \left( E_t U_{t+1}^{1-\gamma} \right)^{\frac{1-\frac{1}{\psi}}{1-\gamma}} \right]^{\frac{1}{1-\frac{1}{\psi}}}, \tag{2}$$

where $c_t$ is the aggregate consumption level, $\beta$ the subjective discount factor, $\gamma$ a risk aversion coefficient, and $\psi$ the EIS. When $\gamma = 1/\psi$, (2) specialize to the common expected utility specification.

As shown in Epstein (1988), Weil (1989), and Epstein and Zin (1989), the stochastic discount factor $M_{t+1}$ is:

$$M_{t+1} = \left[ \beta \left( \frac{c_{t+1}}{c_t} \right)^{-\frac{1}{\psi}} \right]^{\frac{1-\gamma}{1-(1/\psi)}} \left( R_{pt+1} \right)^{\frac{1-\gamma}{1-(1/\psi)} - 1}, \tag{3}$$

where $R_{pt+1}$ is the gross return of the representative agent's portfolio. Let $R_{t+1} = (p_{t+1} + q_{t+1})/p_t$ denote the equity's one-period gross return, where $p_t$ is the price at time t of the equity share (prices are in terms of the time t consumption good). Let $1/R_F$ denote the price of a riskless security, where $R_F - 1$ is the risk free rate. The Euler equations

$$E_t(M_{t+1} R_{t+1}) = 1 \tag{4}$$

$$E_t(M_{t+1}R_F) = 1 \tag{5}$$

allow us to price the equity share and the riskless security (Epstein 1988, Weil 1989 and Epstein and Zin 1989). When $\gamma = 1/\psi$, (4) and (5) specialize to the familiar C-CAPM's Euler equations (Rubinstein 1976 and Lucas 1978).

In equilibrium, the entirety of period t's perishable output is consumed during that period: $c_t = q_t \ \forall t$, and the financial market is cleared, so that the representative agent's portfolio is composed of the equity share: $R_{pt+1} = R_{t+1}$. Then the Euler equations (4) and (5) simplify to:

$$E_t\left\{\left[\beta y_{t+1}^{-1/\psi}\right]^{\frac{1-\gamma}{1-(1/\psi)}} (R_{t+1})^{\frac{1-\gamma}{1-(1/\psi)}}\right\} = 1, \tag{6}$$

$$E_t\left\{\left[\beta y_{t+1}^{-1/\psi}\right]^{\frac{1-\gamma}{1-(1/\psi)}} (R_{t+1})^{\frac{1-\gamma}{1-(1/\psi)}-1} R_F\right\} = 1. \tag{7}$$

The homogeneous price function $p_t = cq_t$, with $c > 0$, solves equation (6). We demonstrate in the appendix that c satisfies to the following condition[2]:

$$\left(\frac{c}{1+c}\right) = \exp\left[-\delta + \left(1-\frac{1}{\psi}\right)\mu + \frac{1}{2}\left(1-\frac{1}{\psi}\right)(1-\gamma)\sigma^2\right], \text{ with } \beta = e^{-\delta}. \tag{8}$$

In equilibrium, the risk free rate and the risk premium are:

$$\ln R_F = \delta + \frac{1}{\psi}\left(\mu + \frac{1}{2}\sigma^2\right) - \frac{1}{2}\gamma\left(1+\frac{1}{\psi}\right)\sigma^2, \tag{9}$$

$$\ln E(R_{t+1}) - \ln R_F = \gamma\sigma^2. \tag{10}$$

According to equation (9), the risk free rate is high when the discount parameter $\delta$ is high; a high interest rate is required to convince investors to save rather than to consume. The risk free rate is also high when the logarithmic expected growth rate[3] $\mu + \sigma^2/2$ is high and it is low when the EIS is high. As the representative consumer feels aversion for intertemporal fluctuations, the desire is to consume more when the expected growth rate is positive. The real interest rate is then pushed upward to restore equilibrium. Finally, the risk free rate is high when volatility of consumption $\sigma^2$ is low. Volatility of consumption $\sigma^2$ captures precautionary saving; the representative consumer is more concerned with low consumption states than he is pleased by high consumption states. The equity premium stated in equation (10) is high when the coefficient of risk aversion and the variance of the growth rate of consumption are high.

Epstein (1988) demonstrates, based on a model similar to ours, (except that his model supposes a stationary economy where the production level $q_t$ is i.i.d.), that the effect on risky asset prices of a variation of the risk aversion parameter depends on the value of the EIS. Risky asset price is an increasing function of the risk aversion coefficient if the EIS is more than one. We demonstrate that this result holds true if the economy is non-stationary.

Solve for c in equation (8) and substitute in the homogenous price function $p_t = cq_t$ to deduce that:

---

[2] See the Appendix for a derivation of equations (8), (9), (10) and (11).
[3] According to (1), the production growth rate follows a lognormal process, therefore, $\ln E(y_{t+1}) = \mu + \sigma^2/2$.

$$p_t = \frac{\exp\left[\mu - \frac{\sigma^2}{2} - \ln E(R_{t+1})\right] q_t}{1 - \exp\left[\mu - \frac{\sigma^2}{2} - \ln E(R_{t+1})\right]}. \quad (11)$$

The derivative $\partial p_t / \partial \gamma$ has the opposite sign to that of $\partial \ln E(R_{t+1}) / \partial \gamma$, namely:

$$\frac{\partial \ln E(R_{t+1})}{\partial \gamma} = \frac{\partial \ln R_F}{\partial \gamma} + \frac{\partial \left(\ln E(R_{t+1}) - \ln R_F\right)}{\partial \gamma}. \quad (12)$$

The consequences of increased risk aversion on equity returns are ambiguous because the risk premium will move in the opposite direction to interest rate changes according to equations (9) and (10). Then, if $\gamma$ rises, the risk free rate may fall sufficiently to induce a drop in expected returns, despite an increase in the risk premium.

From equations (9) and (10) we get the derivative:

$$\frac{\partial \ln E(R_{t+1})}{\partial \gamma} = \frac{\sigma^2}{2}\left(1 - \frac{1}{\psi}\right). \quad (13)$$

This result is similar to that achieved by Epstein (1988) in the case of a stationary economy and can be interpreted in the same way. An increase in risk aversion acts to reduce the certainty equivalent return to saving. "If $\psi < (>)1$, the dominant income (substitution) effect implies reduced (enhanced) present consumption and an increased (reduced) demand for securities" (Epstein 1988, p. 189). Thus, expected equity return is forced to decrease (to increase), and asset price is forced to increase (decrease).

According to equation (13), we may consider that the condition $\psi > 1$ is bound to be verified. This parameter ensures that the substitution effect dominates the wealth effect, so that the representative agent reduces his risky asset demand if he is more risk adverse. If not, financial markets populated by fearful individuals would be characterized by high-priced equities, in comparison with markets populated by confident agents. In this case episodes of financial panic would be characterized by higher asset prices, which has never been observed.

## 4. Conclusion

Is the EIS more or less than one? There is no consensus in the empirical literature about the value of this preference parameter. We can only report that the estimates of the EIS are rather in favour of a low EIS. Nevertheless, more and more authors are developing and estimating asset pricing models based on a high EIS hypothesis, following the long-run risk model of Bansal and Yaron (2004).

By confronting theoretical implications of asset pricing models to actual investor behaviour during episodes of panic, we have shown that the EIS does indeed have to be greater than one, thus confirming the hypotheses of previous authors. If not, the risky asset price could rise when agents are more fearful. As everybody knows, stock prices are lower during episodes of a stock market panic, which is perhaps the strongest proof that EIS is more than one.

# Appendix

**Proof of equation (8):**
Substituting $p_t = cq_t$ and $p_{t+1} = cq_{t+1}$ in $R_{t+1}$, we get:
$$R_{t+1} = \frac{p_{t+1} + q_{t+1}}{p_t} = \frac{cq_{t+1} + q_{t+1}}{cq_t} = \frac{1+c}{c}\frac{q_{t+1}}{q_t} = \frac{1+c}{c} y_{t+1}.$$

Using this result, equation (6) can be written $\left(\frac{c}{1+c}\right) = \beta \left[ E\{y_{t+1}^{1-\gamma}\}\right]^{\frac{1-(1/\psi)}{1-\gamma}}$. Using the property of a normal variable z: $E[e^z] = e^{E[z] + \frac{1}{2}V[z]}$, this last equation can be rearranged to obtain equation (8).

**Proof of equation (9):**
The gross return on equity is proportional to the lognormal dividend growth rate: $R_{t+1} = \frac{1+c}{c} y_{t+1}$. Then, $(\ln R_{t+1}, \ln y_{t+1})$ are jointly normally distributed. Moreover, because the growth rate of dividend and the return on equity are i.i.d., the conditional and unconditional expectations of any function of $y_{t+1}$ and $R_{t+1}$ are the same. Thus, equation (7) can be written in the form: $\beta^{\frac{1-\gamma}{1-(1/\psi)}} \left(\frac{1+c}{c}\right)^{\frac{1-\gamma}{1-(1/\psi)} - 1} E\{y_{t+1}^{-\gamma}\} R_F = 1$. Substituting for (1+c)/c from equation (11) and using the lognormal distribution assumption, we obtain:

$$\exp\left\{-\frac{\delta(1-\gamma)}{1-(1/\psi)}\right\}\left[\exp\left\{-\delta + \left(1-\frac{1}{\psi}\right)\mu + \frac{1}{2}\left(1-\frac{1}{\psi}\right)(1-\gamma)\sigma^2\right\}\right]^{1-\frac{1-\gamma}{1-(1/\psi)}} \exp\left\{-\gamma\mu + \frac{1}{2}\gamma^2\sigma^2\right\} R_F = 1.$$

Taking logs on both sides and simplifying we obtain equation (9).

**Proof of equation (10):**
Given that $(\ln R_{t+1}, \ln y_{t+1})$ are jointly normally distributed, equation (6) can be rearranged:

$$\frac{1-\gamma}{1-(1/\psi)}\left[\ln\beta - \frac{\mu}{\psi}\right] + \frac{1-\gamma}{1-(1/\psi)} E(\ln R_{t+1}) + \frac{1}{2}\left(\frac{1-\gamma}{1-(1/\psi)}\right)^2 V(\ln R_{t+1} - \rho \ln y_{t+1}) = 0$$

Note that $\ln R_{t+1} = \ln\frac{1+c}{c} + \ln y_{t+1}$ to obtain:

$$E(\ln R_{t+1}) = \delta + \frac{\mu}{\psi} - \frac{1}{2}\left(1 - \frac{1}{\psi}\right)(1-\gamma)\sigma^2 \tag{A1}$$

Subtracting (12) from (A1), we find that:
$$E(\ln R_{t+1}) + \frac{1}{2}\sigma^2 - \ln R_F = \gamma\sigma^2 \tag{A2}$$

The lognormal distribution assumption implies that:
$$\ln E(R_{t+1}) = E(\ln R_{t+1}) + \frac{1}{2}V(\ln R_{t+1}), \tag{A3}$$

Substituting (A3) in (A2) results in $\ln E(R_{t+1}) - \ln R_F = \gamma\sigma^2$.

**Proof of equation (11):**
Solve equation (8) for c and substitute in $p_t = cq_t$ to obtain:

$$p_t = \frac{\exp\left[-\delta + \left(1 - \frac{1}{\psi}\right)\mu + \frac{1}{2}\left(1 - \frac{1}{\psi}\right)(1-\gamma)\sigma^2\right] q_t}{1 - \exp\left[-\delta + \left(1 - \frac{1}{\psi}\right)\mu + \frac{1}{2}\left(1 - \frac{1}{\psi}\right)(1-\gamma)\sigma^2\right]}. \quad (A4)$$

Combine (A1) and (A3) to find equation (11).